\documentstyle[11pt,aaspp4]{article}
\newcommand{\kms}{\mbox{ km~s$^{-1}$}}
\newcommand{\kmsM}{\mbox{ km~s$^{-1}$~Mpc$^{-1}$}}
\newcommand{\etal}{et~al.}
\def\PsfigVersion{1.10}
\def\setDriver{\DvipsDriver} 
\ifx\undefined\psfig\else \fi
\let\LaTeXAtSign=\@
\let\@=\relax
\edef\psfigRestoreAt{\catcode`\@=\number\catcode`@\relax}
\catcode`\@=11\relax
\newwrite\@unused
\def\ps@typeout#1{{\let\protect\string\immediate\write\@unused{#1}}}

\def\DvipsDriver{
	\ps@typeout{psfig/tex \PsfigVersion -dvips}
\def\PsfigSpecials{\DvipsSpecials} 	\def\ps@dir{/}
\def\ps@predir{} }
\def\OzTeXDriver{
	\ps@typeout{psfig/tex \PsfigVersion -oztex}
	\def\PsfigSpecials{\OzTeXSpecials}
	\def\ps@dir{:}
	\def\ps@predir{:}
	\catcode`\^^J=5
}

\def\figurepath{./:}

\def\DoPaths#1{\expandafter\EachPath#1\stoplist}
\def\leer{}
\def\EachPath#1:#2\stoplist{
  \ExistsFile{#1}{\SearchedFile}
  \ifx#2\leer
  \else
    \expandafter\EachPath#2\stoplist
  \fi}
\def\ps@dir{/}
\def\ExistsFile#1#2{%
   \openin1=\ps@predir#1\ps@dir#2
   \ifeof1
       \closein1
   \else
       \closein1
        \ifx\ps@founddir\leer
           \edef\ps@founddir{#1}
        \fi
   \fi}
\def\get@dir#1{%
  \def\ps@founddir{}
  \def\SearchedFile{#1}
  \DoPaths\figurepath
}

\def\@nnil{\@nil}
\def\@empty{}
\def\@psdonoop#1\@@#2#3{}
\def\@psdo#1:=#2\do#3{\edef\@psdotmp{#2}\ifx\@psdotmp\@empty \else
    \expandafter\@psdoloop#2,\@nil,\@nil\@@#1{#3}\fi}
\def\@psdoloop#1,#2,#3\@@#4#5{\def#4{#1}\ifx #4\@nnil \else
       #5\def#4{#2}\ifx #4\@nnil \else#5\@ipsdoloop #3\@@#4{#5}\fi\fi}
\def\@ipsdoloop#1,#2\@@#3#4{\def#3{#1}\ifx #3\@nnil 
       \let\@nextwhile=\@psdonoop \else
      #4\relax\let\@nextwhile=\@ipsdoloop\fi\@nextwhile#2\@@#3{#4}}
\def\@tpsdo#1:=#2\do#3{\xdef\@psdotmp{#2}\ifx\@psdotmp\@empty \else
    \@tpsdoloop#2\@nil\@nil\@@#1{#3}\fi}
\def\@tpsdoloop#1#2\@@#3#4{\def#3{#1}\ifx #3\@nnil 
       \let\@nextwhile=\@psdonoop \else
      #4\relax\let\@nextwhile=\@tpsdoloop\fi\@nextwhile#2\@@#3{#4}}
\ifx\undefined\fbox
\newdimen\fboxrule
\newdimen\fboxsep
\newdimen\ps@tempdima
\newbox\ps@tempboxa
\long\def\fbox#1{\leavevmode\setbox\ps@tempboxa\hbox{#1}\ps@tempdima\fboxrule
    \advance\ps@tempdima \fboxsep \advance\ps@tempdima \dp\ps@tempboxa
   \hbox{\lower \ps@tempdima\hbox
  {\vbox{\hrule height \fboxrule
          \hbox{\vrule width \fboxrule \hskip\fboxsep
          \vbox{\vskip\fboxsep \box\ps@tempboxa\vskip\fboxsep}\hskip 
                 \fboxsep\vrule width \fboxrule}
                 \hrule height \fboxrule}}}}
\fi
\newread\ps@stream
\newif\ifnot@eof       
\newif\if@noisy        
\newif\if@atend        
\newif\if@psfile       
{\catcode`\%=12\global\gdef\epsf@start{
\def\epsf@PS{PS}
\def\epsf@getbb#1{%
\openin\ps@stream=\ps@predir#1
\ifeof\ps@stream\ps@typeout{Error, File #1 not found}\else
   {\not@eoftrue \chardef\other=12
    \def\do##1{\catcode`##1=\other}\dospecials \catcode`\ =10
    \loop
       \if@psfile
	  \read\ps@stream to \epsf@fileline
       \else{
	  \obeyspaces
          \read\ps@stream to \epsf@tmp\global\let\epsf@fileline\epsf@tmp}
       \fi
       \ifeof\ps@stream\not@eoffalse\else
       \if@psfile\else
       \expandafter\epsf@test\epsf@fileline:. \\%
       \fi
          \expandafter\epsf@aux\epsf@fileline:. \\%
       \fi
   \ifnot@eof\repeat
   }\closein\ps@stream\fi}%
\long\def\epsf@test#1#2#3:#4\\{\def\epsf@testit{#1#2}
			\ifx\epsf@testit\epsf@start\else
\ps@typeout{Warning! File does not start with `\epsf@start'.  It may not be a PostScript file.}
			\fi
			\@psfiletrue} 
{\catcode`\%=12\global\let\epsf@percent=
\long\def\epsf@aux#1#2:#3\\{\ifx#1\epsf@percent
   \def\epsf@testit{#2}\ifx\epsf@testit\epsf@bblit
	\@atendfalse
        \epsf@atend #3 . \\%
	\if@atend	
	   \if@verbose{
		\ps@typeout{psfig: found `(atend)'; continuing search}
	   }\fi
        \else
        \epsf@grab #3 . . . \\%
        \not@eoffalse
        \global\no@bbfalse
        \fi
   \fi\fi}%
\def\epsf@grab #1 #2 #3 #4 #5\\{%
   \global\def\epsf@llx{#1}\ifx\epsf@llx\empty
      \epsf@grab #2 #3 #4 #5 .\\\else
   \global\def\epsf@lly{#2}%
   \global\def\epsf@urx{#3}\global\def\epsf@ury{#4}\fi}%
\def\epsf@atendlit{(atend)} 
\def\epsf@atend #1 #2 #3\\{%
   \def\epsf@tmp{#1}\ifx\epsf@tmp\empty
      \epsf@atend #2 #3 .\\\else
   \ifx\epsf@tmp\epsf@atendlit\@atendtrue\fi\fi}

\chardef\psletter = 11 
\chardef\other = 12

\newif \ifdebug 
\newif\ifc@mpute 
\c@mputetrue 

\let\then = \relax
\def\r@dian{pt }
\let\r@dians = \r@dian
\let\dimensionless@nit = \r@dian
\let\dimensionless@nits = \dimensionless@nit
\def\internal@nit{sp }
\let\internal@nits = \internal@nit
\newif\ifstillc@nverging
\def \Mess@ge #1{\ifdebug \then \message {#1} \fi}

{ 
	\catcode `\@ = \psletter
	\gdef \nodimen {\expandafter \n@dimen \the \dimen}
	\gdef \term #1 #2 #3%
	       {\edef \t@ {\the #1}
		\edef \t@@ {\expandafter \n@dimen \the #2\r@dian}%
		\t@rm {\t@} {\t@@} {#3}%
	       }
	\gdef \t@rm #1 #2 #3%
	       {{%
		\count 0 = 0
		\dimen 0 = 1 \dimensionless@nit
		\dimen 2 = #2\relax
		\Mess@ge {Calculating term #1 of \nodimen 2}%
		\loop
		\ifnum	\count 0 < #1
		\then	\advance \count 0 by 1
			\Mess@ge {Iteration \the \count 0 \space}%
			\Multiply \dimen 0 by {\dimen 2}%
			\Mess@ge {After multiplication, term = \nodimen 0}%
			\Divide \dimen 0 by {\count 0}%
			\Mess@ge {After division, term = \nodimen 0}%
		\repeat
		\Mess@ge {Final value for term #1 of 
				\nodimen 2 \space is \nodimen 0}%
		\xdef \Term {#3 = \nodimen 0 \r@dians}%
		\aftergroup \Term
	       }}
	\catcode `\p = \other
	\catcode `\t = \other
	\gdef \n@dimen #1pt{#1} 
}

\def \Divide #1by #2{\divide #1 by #2} 

\def \Multiply #1by #2
       {{
	\count 0 = #1\relax
	\count 2 = #2\relax
	\count 4 = 65536
	\Mess@ge {Before scaling, count 0 = \the \count 0 \space and
			count 2 = \the \count 2}%
	\ifnum	\count 0 > 32767 
	\then	\divide \count 0 by 4
		\divide \count 4 by 4
	\else	\ifnum	\count 0 < -32767
		\then	\divide \count 0 by 4
			\divide \count 4 by 4
		\else
		\fi
	\fi
	\ifnum	\count 2 > 32767 
	\then	\divide \count 2 by 4
		\divide \count 4 by 4
	\else	\ifnum	\count 2 < -32767
		\then	\divide \count 2 by 4
			\divide \count 4 by 4
		\else
		\fi
	\fi
	\multiply \count 0 by \count 2
	\divide \count 0 by \count 4
	\xdef \product {#1 = \the \count 0 \internal@nits}%
	\aftergroup \product
       }}

\def\r@duce{\ifdim\dimen0 > 90\r@dian \then   
		\multiply\dimen0 by -1
		\advance\dimen0 by 180\r@dian
		\r@duce
	    \else \ifdim\dimen0 < -90\r@dian \then  
		\advance\dimen0 by 360\r@dian
		\r@duce
		\fi
	    \fi}

\def\Sine#1%
       {{%
	\dimen 0 = #1 \r@dian
	\r@duce
	\ifdim\dimen0 = -90\r@dian \then
	   \dimen4 = -1\r@dian
	   \c@mputefalse
	\fi
	\ifdim\dimen0 = 90\r@dian \then
	   \dimen4 = 1\r@dian
	   \c@mputefalse
	\fi
	\ifdim\dimen0 = 0\r@dian \then
	   \dimen4 = 0\r@dian
	   \c@mputefalse
	\fi
	\ifc@mpute \then
		\divide\dimen0 by 180
		\dimen0=3.141592654\dimen0
		\dimen 2 = 3.1415926535897963\r@dian 
		\divide\dimen 2 by 2 
		\Mess@ge {Sin: calculating Sin of \nodimen 0}%
		\count 0 = 1 
		\dimen 2 = 1 \r@dian 
		\dimen 4 = 0 \r@dian 
		\loop
			\ifnum	\dimen 2 = 0 
			\then	\stillc@nvergingfalse 
			\else	\stillc@nvergingtrue
			\fi
			\ifstillc@nverging 
			\then	\term {\count 0} {\dimen 0} {\dimen 2}%
				\advance \count 0 by 2
				\count 2 = \count 0
				\divide \count 2 by 2
				\ifodd	\count 2 
				\then	\advance \dimen 4 by \dimen 2
				\else	\advance \dimen 4 by -\dimen 2
				\fi
		\repeat
	\fi		
			\xdef \sine {\nodimen 4}%
       }}

\def\Cosine#1{\ifx\sine\UnDefined\edef\Savesine{\relax}\else
		             \edef\Savesine{\sine}\fi
	{\dimen0=#1\r@dian\advance\dimen0 by 90\r@dian
	 \Sine{\nodimen 0}
	 \xdef\cosine{\sine}
	 \xdef\sine{\Savesine}}}	      

\def\psdraft{
	\def\@psdraft{0}
}
\def\psfull{
	\def\@psdraft{100}
}

\psfull

\newif\if@scalefirst
\def\psscalefirst{\@scalefirsttrue}
\def\psrotatefirst{\@scalefirstfalse}
\psrotatefirst

\newif\if@draftbox
\def\psnodraftbox{
	\@draftboxfalse
}
\def\psdraftbox{
	\@draftboxtrue
}
\@draftboxtrue

\newif\if@prologfile
\newif\if@postlogfile
\def\pssilent{
	\@noisyfalse
}
\def\psnoisy{
	\@noisytrue
}
\psnoisy
\newif\if@bbllx
\newif\if@bblly
\newif\if@bburx
\newif\if@bbury
\newif\if@height
\newif\if@width
\newif\if@rheight
\newif\if@rwidth
\newif\if@angle
\newif\if@clip
\newif\if@verbose
\def\@p@@sclip#1{\@cliptrue}
\newif\if@decmpr
\def\@p@@sfigure#1{\def\@p@sfile{null}\def\@p@sbbfile{null}\@decmprfalse
   \openin1=\ps@predir#1
   \ifeof1
	\closein1
	\get@dir{#1}
	\ifx\ps@founddir\leer
		\openin1=\ps@predir#1.bb
		\ifeof1
			\closein1
			\get@dir{#1.bb}
			\ifx\ps@founddir\leer
				\ps@typeout{Can't find #1 in \figurepath}
			\else
				\@decmprtrue
				\def\@p@sfile{\ps@founddir\ps@dir#1}
				\def\@p@sbbfile{\ps@founddir\ps@dir#1.bb}
			\fi
		\else
			\closein1
			\@decmprtrue
			\def\@p@sfile{#1}
			\def\@p@sbbfile{#1.bb}
		\fi
	\else
		\def\@p@sfile{\ps@founddir\ps@dir#1}
		\def\@p@sbbfile{\ps@founddir\ps@dir#1}
	\fi
   \else
	\closein1
	\def\@p@sfile{#1}
	\def\@p@sbbfile{#1}
   \fi
}
\def\@p@@sfile#1{\@p@@sfigure{#1}}
\def\@p@@sbbllx#1{
		\@bbllxtrue
		\dimen100=#1
		\edef\@p@sbbllx{\number\dimen100}
}
\def\@p@@sbblly#1{
		\@bbllytrue
		\dimen100=#1
		\edef\@p@sbblly{\number\dimen100}
}
\def\@p@@sbburx#1{
		\@bburxtrue
		\dimen100=#1
		\edef\@p@sbburx{\number\dimen100}
}
\def\@p@@sbbury#1{
		\@bburytrue
		\dimen100=#1
		\edef\@p@sbbury{\number\dimen100}
}
\def\@p@@sheight#1{
		\@heighttrue
		\dimen100=#1
   		\edef\@p@sheight{\number\dimen100}
}
\def\@p@@swidth#1{
		\@widthtrue
		\dimen100=#1
		\edef\@p@swidth{\number\dimen100}
}
\def\@p@@srheight#1{
		\@rheighttrue
		\dimen100=#1
		\edef\@p@srheight{\number\dimen100}
}
\def\@p@@srwidth#1{
		\@rwidthtrue
		\dimen100=#1
		\edef\@p@srwidth{\number\dimen100}
}
\def\@p@@sangle#1{
		\@angletrue
		\edef\@p@sangle{#1} 
}
\def\@p@@ssilent#1{ 
		\@verbosefalse
}
\def\@p@@sprolog#1{\@prologfiletrue\def\@prologfileval{#1}}
\def\@p@@spostlog#1{\@postlogfiletrue\def\@postlogfileval{#1}}
\def\@cs@name#1{\csname #1\endcsname}
\def\@setparms#1=#2,{\@cs@name{@p@@s#1}{#2}}
\def\ps@init@parms{
		\@bbllxfalse \@bbllyfalse
		\@bburxfalse \@bburyfalse
		\@heightfalse \@widthfalse
		\@rheightfalse \@rwidthfalse
		\def\@p@sbbllx{}\def\@p@sbblly{}
		\def\@p@sbburx{}\def\@p@sbbury{}
		\def\@p@sheight{}\def\@p@swidth{}
		\def\@p@srheight{}\def\@p@srwidth{}
		\def\@p@sangle{0}
		\def\@p@sfile{} \def\@p@sbbfile{}
		\def\@p@scost{10}
		\def\@sc{}
		\@prologfilefalse
		\@postlogfilefalse
		\@clipfalse
		\if@noisy
			\@verbosetrue
		\else
			\@verbosefalse
		\fi
}
\def\parse@ps@parms#1{
	 	\@psdo\@psfiga:=#1\do
		   {\expandafter\@setparms\@psfiga,}}
\newif\ifno@bb
\def\bb@missing{
	\if@verbose{
		\ps@typeout{psfig: searching \@p@sbbfile \space  for bounding box}
	}\fi
	\no@bbtrue
	\epsf@getbb{\@p@sbbfile}
        \ifno@bb \else \bb@cull\epsf@llx\epsf@lly\epsf@urx\epsf@ury\fi
}	
\def\bb@cull#1#2#3#4{
	\dimen100=#1 bp\edef\@p@sbbllx{\number\dimen100}
	\dimen100=#2 bp\edef\@p@sbblly{\number\dimen100}
	\dimen100=#3 bp\edef\@p@sbburx{\number\dimen100}
	\dimen100=#4 bp\edef\@p@sbbury{\number\dimen100}
	\no@bbfalse
}
\newdimen\p@intvaluex
\newdimen\p@intvaluey
\def\rotate@#1#2{{\dimen0=#1 sp\dimen1=#2 sp
		  \global\p@intvaluex=\cosine\dimen0
		  \dimen3=\sine\dimen1
		  \global\advance\p@intvaluex by -\dimen3
		  \global\p@intvaluey=\sine\dimen0
		  \dimen3=\cosine\dimen1
		  \global\advance\p@intvaluey by \dimen3
		  }}
\def\compute@bb{
		\no@bbfalse
		\if@bbllx \else \no@bbtrue \fi
		\if@bblly \else \no@bbtrue \fi
		\if@bburx \else \no@bbtrue \fi
		\if@bbury \else \no@bbtrue \fi
		\ifno@bb \bb@missing \fi
		\ifno@bb \ps@typeout{FATAL ERROR: no bb supplied or found}
			\no-bb-error
		\fi
		\count203=\@p@sbburx
		\count204=\@p@sbbury
		\advance\count203 by -\@p@sbbllx
		\advance\count204 by -\@p@sbblly
		\edef\ps@bbw{\number\count203}
		\edef\ps@bbh{\number\count204}
		\if@angle 
			\Sine{\@p@sangle}\Cosine{\@p@sangle}
	        	{\dimen100=\maxdimen\xdef\r@p@sbbllx{\number\dimen100}
					    \xdef\r@p@sbblly{\number\dimen100}
			                    \xdef\r@p@sbburx{-\number\dimen100}
					    \xdef\r@p@sbbury{-\number\dimen100}}
                        \def\minmaxtest{
			   \ifnum\number\p@intvaluex<\r@p@sbbllx
			      \xdef\r@p@sbbllx{\number\p@intvaluex}\fi
			   \ifnum\number\p@intvaluex>\r@p@sbburx
			      \xdef\r@p@sbburx{\number\p@intvaluex}\fi
			   \ifnum\number\p@intvaluey<\r@p@sbblly
			      \xdef\r@p@sbblly{\number\p@intvaluey}\fi
			   \ifnum\number\p@intvaluey>\r@p@sbbury
			      \xdef\r@p@sbbury{\number\p@intvaluey}\fi
			   }
			\rotate@{\@p@sbbllx}{\@p@sbblly}
			\minmaxtest
			\rotate@{\@p@sbbllx}{\@p@sbbury}
			\minmaxtest
			\rotate@{\@p@sbburx}{\@p@sbblly}
			\minmaxtest
			\rotate@{\@p@sbburx}{\@p@sbbury}
			\minmaxtest
			\edef\@p@sbbllx{\r@p@sbbllx}\edef\@p@sbblly{\r@p@sbblly}
			\edef\@p@sbburx{\r@p@sbburx}\edef\@p@sbbury{\r@p@sbbury}
		\fi
		\count203=\@p@sbburx
		\count204=\@p@sbbury
		\advance\count203 by -\@p@sbbllx
		\advance\count204 by -\@p@sbblly
		\edef\@bbw{\number\count203}
		\edef\@bbh{\number\count204}
}
\def\in@hundreds#1#2#3{\count240=#2 \count241=#3
		     \count100=\count240	
		     \divide\count100 by \count241
		     \count101=\count100
		     \multiply\count101 by \count241
		     \advance\count240 by -\count101
		     \multiply\count240 by 10
		     \count101=\count240	
		     \divide\count101 by \count241
		     \count102=\count101
		     \multiply\count102 by \count241
		     \advance\count240 by -\count102
		     \multiply\count240 by 10
		     \count102=\count240	
		     \divide\count102 by \count241
		     \count200=#1\count205=0
		     \count201=\count200
			\multiply\count201 by \count100
		 	\advance\count205 by \count201
		     \count201=\count200
			\divide\count201 by 10
			\multiply\count201 by \count101
			\advance\count205 by \count201
		     \count201=\count200
			\divide\count201 by 100
			\multiply\count201 by \count102
			\advance\count205 by \count201
		     \edef\@result{\number\count205}
}
\def\compute@wfromh{
		\in@hundreds{\@p@sheight}{\@bbw}{\@bbh}
		\edef\@p@swidth{\@result}
}
\def\compute@hfromw{
	        \in@hundreds{\@p@swidth}{\@bbh}{\@bbw}
		\edef\@p@sheight{\@result}
}
\def\compute@handw{
		\if@height 
			\if@width
			\else
				\compute@wfromh
			\fi
		\else 
			\if@width
				\compute@hfromw
			\else
				\edef\@p@sheight{\@bbh}
				\edef\@p@swidth{\@bbw}
			\fi
		\fi
}
\def\compute@resv{
		\if@rheight \else \edef\@p@srheight{\@p@sheight} \fi
		\if@rwidth \else \edef\@p@srwidth{\@p@swidth} \fi
}
\def\compute@sizes{
	\compute@bb
	\if@scalefirst\if@angle
	\if@width
	   \in@hundreds{\@p@swidth}{\@bbw}{\ps@bbw}
	   \edef\@p@swidth{\@result}
	\fi
	\if@height
	   \in@hundreds{\@p@sheight}{\@bbh}{\ps@bbh}
	   \edef\@p@sheight{\@result}
	\fi
	\fi\fi
	\compute@handw
	\compute@resv}
\def\OzTeXSpecials{
	\special{empty.ps /@isp {true} def}
	\special{empty.ps \@p@swidth \space \@p@sheight \space
			\@p@sbbllx \space \@p@sbblly \space
			\@p@sbburx \space \@p@sbbury \space
			startTexFig \space }
	\if@clip{
		\if@verbose{
			\ps@typeout{(clip)}
		}\fi
		\special{empty.ps doclip \space }
	}\fi
	\if@angle{
		\if@verbose{
			\ps@typeout{(rotate)}
		}\fi
		\special {empty.ps \@p@sangle \space rotate \space} 
	}\fi
	\if@prologfile
	    \special{\@prologfileval \space } \fi
	\if@decmpr{
		\if@verbose{
			\ps@typeout{psfig: Compression not available
			in OzTeX version \space }
		}\fi
	}\else{
		\if@verbose{
			\ps@typeout{psfig: including \@p@sfile \space }
		}\fi
		\special{epsf=\ps@predir\@p@sfile \space }
	}\fi
	\if@postlogfile
	    \special{\@postlogfileval \space } \fi
	\special{empty.ps /@isp {false} def}
}
\def\DvipsSpecials{
	\special{ps::[begin] 	\@p@swidth \space \@p@sheight \space
			\@p@sbbllx \space \@p@sbblly \space
			\@p@sbburx \space \@p@sbbury \space
			startTexFig \space }
	\if@clip{
		\if@verbose{
			\ps@typeout{(clip)}
		}\fi
		\special{ps:: doclip \space }
	}\fi
	\if@angle
		\if@verbose{
			\ps@typeout{(clip)}
		}\fi
		\special {ps:: \@p@sangle \space rotate \space} 
	\fi
	\if@prologfile
	    \special{ps: plotfile \@prologfileval \space } \fi
	\if@decmpr{
		\if@verbose{
			\ps@typeout{psfig: including \@p@sfile.Z \space }
		}\fi
		\special{ps: plotfile "`zcat \@p@sfile.Z" \space }
	}\else{
		\if@verbose{
			\ps@typeout{psfig: including \@p@sfile \space }
		}\fi
		\special{ps: plotfile \@p@sfile \space }
	}\fi
	\if@postlogfile
	    \special{ps: plotfile \@postlogfileval \space } \fi
	\special{ps::[end] endTexFig \space }
}
\def\psfig#1{\vbox {
	\ps@init@parms
	\parse@ps@parms{#1}
	\compute@sizes
	\ifnum\@p@scost<\@psdraft{
		\PsfigSpecials 
		\vbox to \@p@srheight sp{
			\hbox to \@p@srwidth sp{
				\hss
			}
		\vss
		}
	}\else{
		\if@draftbox{		
			\hbox{\fbox{\vbox to \@p@srheight sp{
			\vss
			\hbox to \@p@srwidth sp{ \hss 
			 \hss }
			\vss
			}}}
		}\else{
			\vbox to \@p@srheight sp{
			\vss
			\hbox to \@p@srwidth sp{\hss}
			\vss
			}
		}\fi

	}\fi
}}
\psfigRestoreAt
\setDriver
\let\@=\LaTeXAtSign

\begin{document}

\title {Multi-Frequency Analysis of the New Wide-Separation Gravitational Lens
Candidate RX\,J0921+4529\footnote{
             Based on Observations made with the NASA/ESA
             Hubble Space Telescope, obtained at the Space Telescope
             Science Institute, which is operated by AURA, Inc., 
             under NASA contract NAS 5-26555.}$^,$\footnote{Based on observations
             obtained at the Multiple Mirror Telescope
             Observatory, a facility operated jointly by the University of Arizona and the
             Smithsonian Institution.} 
             }

\author{J. A. Mu\~noz\altaffilmark{3,4},
         E. E. Falco\altaffilmark{3}, 
         C. S. Kochanek\altaffilmark{3}, 
         J. Leh\'ar\altaffilmark{3},
         B. A. McLeod\altaffilmark{3}, 
         B. R. McNamara\altaffilmark{3},
         A. A. Vikhlinin\altaffilmark{3},
         C. D. Impey\altaffilmark{5}, 
         H.-W. Rix\altaffilmark{6},
         C. R. Keeton\altaffilmark{5}, 
         C. Y. Peng\altaffilmark{5}
         and C. R. Mullis\altaffilmark{7}}

\altaffiltext{3}{Harvard-Smithsonian Center for Astrophysics, Cambridge, 
MA 02138, USA}
\altaffiltext{4}{Instituto de Astrof\'{\i}sica de Canarias, E-38200 La Laguna, Tenerife, Spain}
\altaffiltext{5}{Steward Observatory, University of Arizona, Tucson, AZ 85721, USA}
\altaffiltext{6}{Max Planck Institut f\"ur Astrophysik, Heidelberg, D-69117, Germany}
\altaffiltext{7}{Institute for Astronomy, University of Hawaii, Honolulu, HI 96822, USA}

\begin{abstract}
We report the discovery of a new two-image gravitational lens 
candidate. The system RX\,J0921+4529 contains two $z_s=1.66$ 
quasars separated by
6\farcs93 with an H band magnitude difference of
$\Delta m=1.39$. The HST NIC2 H band images
reveal an H=18.2 spiral ga\-laxy between the quasar images,
which is probably a member of a $z_l=0.32$ X-ray
cluster centered on the field. 
We detect an extended
source near the fainter quasar image but not in the
brighter image. If this extended source is the host
galaxy of the fainter quasar, then the system is a binary
quasar rather than a gravitational lens. 
VLA observations at 3.6\,cm reveal emission from the
lens galaxy at the flux level of 1\,mJy
and a marginal detection of the brighter quasar.

\end{abstract}
 

\section{Introduction}

\def\dt{\Delta\theta}
\def\dtm{\Delta\theta_{max}}

Gravitational lenses produced by isolated galaxies should have a mean
image separation of 1\farcs5, closely matching the mean separation
observed in the current sample of over 60 gravitational lenses
(e.g. Falco, Kochanek \& Mu\~noz 1998).  Surveys for lenses 
(JVAS, Patnaik \etal\ 1992; MG-VLA, Burke \etal\ 1993;
CLASS, Jackson \etal\ 1998;
see also http://cfa-www.harvard.edu/castles) have demonstrated that
gravitational lenses with image separations larger than 3\farcs0 are rare.
In order of increasing separation we know of HE~1104-1805
($\dt=3\farcs19$), MG~2016+112 ($\dt=3\farcs26$) and Q~0957+561
($\dt=6\farcs26$).\footnote{ These are the standard separations
introduced by Kochanek et~al.\ (2000) which correct for shear
effects. Five lenses have {\it maximum} separations larger than
3\farcs0.  In order of increasing separation, these are HE~1104-1805
($\dtm=3\farcs19$), RX\,J0911+0551 ($\dtm=3\farcs25$), HST~14176+5226
($\dtm=3\farcs28$), MG~2016+112 ($\dtm=3\farcs88$), and Q~0957+561
($\dtm=6\farcs17$).  All five of these systems have massive early-type
galaxies as their primary lens.}  All three of these systems have
massive early-type galaxies as their primary lens (Kochanek et~al.\
2000).

The overwhelming dominance of the lens population by galaxies is a
consequence of the condensation of the baryons in forming galaxies.
Predictions of the distribution of image separations based on the
density and masses of dark matter halos, either from Press-Schechter
models or simulations, catastrophically fail to match the observed
separation distribution of lenses at both high and low image
separations.  The large separation lenses correspond to lenses
produced by groups and small clusters, and Keeton (1998) has shown
that we expect a strong enhancement in lensing by galaxies compared to
groups and small clusters because the ``cooled'' baryons of the
galaxies make them relatively more efficient lenses.  The absolute
numbers of wide separation lenses are in good agreement with
predictions for power spectra normalized to produce the observed
density of clusters (Narayan \& White 1988, Cen \etal\ 1994, Kochanek
1995, Maoz \etal\ 1997, Wambsganss, Cen \& Ostriker 1998).

Discussions of wide separation lenses were confused for many years by the
presence of a larger population of binary quasars with separations of
3\farcs0 to 10\farcs0.  Although the spectral similarities of some binary
quasars remains a puzzle (e.g. Michalitsianos \etal\ 1997,
Peng \etal\ 1999, Mortlock \etal\ 1999), 
Mu\~noz et~al.\ (1998) and Kochanek, Falco \& Mu\~noz (1999) demonstrated that
most of the so-called ``dark lenses'' were binary quasars.  Moreover, 
Kochanek \etal\ (1999) provided a quantitative explanation of their
separations and abundances.  We are now confident that wide separation
lenses should be associated with visible groups or clusters of galaxies
whose higher masses explain the larger separations. 

In this paper we present a new example, RX\,J0921+4529. Its angular
separation of 6\farcs93 would make it the lensed quasar with the
largest angular separation, if lensing were to be confirmed. The
detection of a galaxy lying between the two quasars and the presence
of an X-ray galaxy cluster to explain the wide separation of the images,
make this system a good gravitational lens candidate.  In \S2 we
discuss the discovery of RX\,J0921+4529 and subsequent observations to
measure its properties.  In \S3 we present simple lens models and
discuss their consequences.

\section{Observations}

RX\,J0921+4529 was detected as an extended X-ray source in a 160~deg$^2$ 
ROSAT survey for  high redshift galaxy clusters (Vikhlinin et~al.\ 1998).
The X-ray image was significantly (6-$\sigma$) broader than
the local ROSAT Point Response Function, and the  X-ray surface brightness 
distribution is well fit by a $\beta$-model,
$S(r)=S_0/(1+r^2/r_c^2)^{3\beta-0.5}$ with a $\beta$ fixed at 0.7 and $r_c=26\arcsec$.
The X-ray flux is 2.39 $10^{-13}$ erg/s/cm$^2$ in the 0.5-2 keV band.
At the cluster redshift ($z_l\simeq0.32$), this corresponds to L$_X$=1.0 $10^{44}$ erg/s
(H$_0$=50\kmsM), or a
temperature of 3.7 keV (derived from the L$_X$-T relation).
Optical images were obtained with the Fred Lawrence Whipple
Observatory (FLWO) 1.2m telescope in the R band
to search for a cluster corresponding to the X-ray source.  As illustrated
in Figure~1, there is a significant excess of galaxies in the field, centered
on the X-ray source.  The optical images also revealed two blue point sources 
(labeled A and B in Figure~1), which appeared to be promising
quasar candidates.

We obtained spectra of the point sources (A and B) with the MMT and the Blue
Channel spectrograph. The spectra had integration times of 3000s and
3600s respectively, and they covered a wavelength range of
3230--8800\AA, with a resolution of 1.96 \AA\ ${\rm pixel}^{-1}$ and
an effective resolution (FWHM) of 7 \AA.  Both A and B are $z_s=1.66$
quasars (see Figure~2), with a velocity difference of $|\Delta v| \leq
1500 \kms$ depending on the wavelength range used for the spectral
cross-correlation.  Thus, RX\,J0921+4529A/B are either a gravitational
lens or a binary quasar.  For a lens with a time delay $\sim 100$
days, a velocity difference of $|\Delta v|\sim1500\kms$ 
could be
created by quasar variability coupled with a long time delay (see
Small, Sargent \& Steidel 1997).  Note that the continuum of the B
spectrum is flatter than that of the A spectrum.  In addition to the
two quasars, spectra were obtained for 5 of the nearby galaxies (see
Table~2). Three of them were acquired at the University of Hawaii 2.2m telescope.
All five were found to have redshifts of $z_l\simeq0.32$,
confirming the presence of a cluster and a favorable geometry for
producing a wide separation lens.  We failed to measure the
redshift of the central, putative lens galaxy G, and we will assume it is 
a member of the cluster. 

We obtained HST observations of RX\,J0921+4529, as part of the CASTLES
(CfA/Arizona Space Telescope Lens Survey) project.  The target was
observed for 2560\,s using the NICMOS/NIC2 camera with the F160W
filter (H-band), and for 1800\,s using the WFPC2 camera with the
F814W filter (I-band).  The data were reduced and analyzed using the
standard procedures described in Leh\'ar et~al.\ (2000), see also McLeod (1997).
In the H-band image the A and B quasars are separated by
$6\farcs929\pm0\farcs006$ and have H-band magnitudes of 16.90$\pm$0.03
and 18.29$\pm$0.04 respectively.  Figure~3 shows the surrounding field
from the NICMOS H-band image.  Lying between the two quasars is an
H=$18.2\pm0.4$ (I$-$H=$1.9\pm0.4$) galaxy with apparent spiral structure (see Figure~4 and
Table~1).  All other wide separation lenses have a massive early-type
galaxy which dominates the lensing effects.  In the residual image
found after subtracting the two quasars from the NIC2 H-band image
(see Figure~4), there appears to be an extended, faint object B$'$
near the fainter quasar B.  In our best models there is a small 
but poorly determined shift
($0\farcs3\pm0\farcs2$) between B$'$ and B, suggesting that B$'$ is a
faint cluster galaxy.  However B$'$ is undetected in the I-band image
and we cannot rule out that it corresponds to the host galaxy of the B
quasar.  No comparable extended object is seen near the A quasar.  As
we discuss in \S3, if B$'$ is the host galaxy of quasar B, the absence
of a host galaxy for quasar A means that the system is a perversely
located binary quasar rather than a gravitational lens.

We cataloged the galaxies in the HST images using the SExtractor
package (Bertin \& Arnouts 1996).  The software classified the optical
objects, and computed total magnitudes using Kron-type automatic
apertures.  Colors were computed using fixed apertures scaled to the
F814W size of each object.  Table~2 lists all the galaxies detected by
HST within 20\arcsec\ of the lens, after visually confirming the
SExtractor classification.  The galaxies were labeled G\# in order of
decreasing I-band brightness, or H\# for those NICMOS galaxies 
with no I-band detection.  The large number of galaxies detected
corroborates the presence of the cluster of galaxies.  Figure~5 shows
HST colors and magnitudes for the nearby galaxies, compared to
evolutionary tracks for various spectrophotometric models (Bruzual \&
Charlot 1993; 2000, in preparation).  The putative lens galaxy G and
its neighbors are photometrically consistent with the $z=0.32$ cluster
redshift.  Galaxy G appears to have spiral structure (see Figure 4)
which our image fitting software is not designed to model.
We modeled the galaxy (Table 1) as a de Vaucouluers bulge 
plus an exponential disk, but the residuals are still large.  
Because the galaxies in this field are larger than most
in the CASTLES target fields, we have increased the color apertures to
a fixed 4\arcsec diameter in all cases.

Finally, we obtained a radio map of RX\,J0921+4529,
using the VLA C~configuration at 3.6\,cm. 
The VLA observations were made on 1998.12.10, with 15 minutes on target.
The beam FWHM was 3\arcsec\ and the detection limit was 0.05 mJy/beam. The
cluster galaxy G2 was clearly detected, so we aligned the optical and radio 
images using it. In the VLA map (see Figure~6), both G2 and the lens galaxy 
G have 8.5~GHz fluxes of $\simeq 1.0 \pm 0.1 $ mJy.  There may be a marginal 
detection of A, at $\simeq 0.1 \pm 0.05$ mJy.   If A is a $0.1$~mJy source,
then we would expect B to be a $\sim 0.03$~mJy source, so a deeper radio
image could further confirm the lensing nature of the system.

\section{Discussion}

With a separation of 6\farcs93, RX\,J0921+4529 is the
widest angular separation lens found without searching directly
behind a rich cluster.  Almost all the data support the lens
interpretation, as we see 2 images of a $z_s=1.66$ lensed quasar with
similar spectra and a modest magnitude difference.  There is
a foreground galaxy located between the quasars to provide
the high surface density region needed to produce two images,
and the galaxy is almost certainly a member of the $z=0.32$
X-ray cluster which provides the additional mass needed to
produce the large separation.

The large angular separation of RX\,J0921+4529 should be produced by 
the combination of the main lens galaxy lying between the two 
quasar images and the presence of 
the galaxy cluster. We estimated the local tidal shear from 
nearby galaxies by assuming
that each is embedded in a singular isothermal sphere halo,
with the same mass-to-light ratio as the lens galaxy G.
We assumed an Einstein ring radius of $b\sim1\arcsec$ for
the lens galaxy, and scaled the other galaxy ring sizes to the
SExtractor I-filter luminosity, with $b\propto\sqrt{L}$ (see Table~2).
A total $\gamma_T\simeq0.17$ in the direction PA=$-36\arcdeg$ is
obtained by taking the tensor sum of the individual contributions.
Because the HST field is not big enough to include the full cluster
this shear estimate represents only a preliminary value. 
Chandra or XMM imaging and deeper wide-field optical images
could clarify the cluster properties
and its contribution to the lensing gravitational potential.

The simplicity of the lens geometry means that we find a wide range of
lens models consistent with our current data.  However with the extant
data, we cannot determine the position of the cluster center, and we
cannot easily include the cluster in the lens models.  Thus, our
objective in this section is only to illustrate the importance of the
interpretation of B$'$ as either a small cluster galaxy or the host of
the B quasar. 
For now we fit the system only with a singular isothermal sphere (SIS) 
and an external shear. Given the lens galaxy and quasar positions, and the quasar
flux ratio (see Table~1) this lens model needs an Einstein radius
$b=3\farcs3$, and an external shear $\gamma=0.047$ orientated in the
direction PA$=33\arcdeg$.  The leading image is the fainter B quasar
and the time delay in a flat ($\Omega=1$) cosmology is $\Delta t
=93\,h^{-1}$ days\footnote{A singular isothermal ellipsoid (SIE) with a fixed
external shear to the observed values (see above) also fits the data
with the lens parameters $b=3\farcs3$, e=0.48 and $\theta_e=51\arcdeg$.
For this model the time delay is $\Delta t =118\,h^{-1}$ days.}. 
In Figure~7 we show both the source and lens
planes.  The source plane includes an image of the lens galaxy G
to orient the figure, a point source for the quasar and its host
galaxy with a de Vaucouleurs profile.  In the image plane we
see the 2 images of the quasar which exactly fit the observed
positions and fluxes. The small arc from the host galaxy near quasar B
has a flux matching that of B$'$.  However, if B$'$ were a host galaxy,
we would see the larger, brighter and undetected arc seen near
image A.

We are reasonably certain that the extended object B$'$ near the B
quasar is a faint cluster galaxy rather than the host galaxy of quasar
B.  If, however, it is the host galaxy, the absence of a larger
host image near quasar A means that the system is a particularly
perverse example of a binary quasar lying behind a foreground X-ray
cluster.  Deeper HST images would determine more accurately the B$'$
position and would clarify the nature of the object.  Deeper radio
images may also resolve the problem if our marginal detection of the A
quasar at 8.5~GHz is real.

\acknowledgements
Acknowledgements:
Support for the CASTLES project was provided by NASA through grant
numbers GO-7495 and GO-7887 from the Space Telescope Science
Institute which is operated by the Association of Universities for
Research in Astronomy, Inc. under NASA contract NAS 5-26555. 
This research was supported in part by
the Smithsonian Institution. CSK is also supported by NASA grant NAG5-4062.

\begin{figure}
\centerline{\psfig{figure=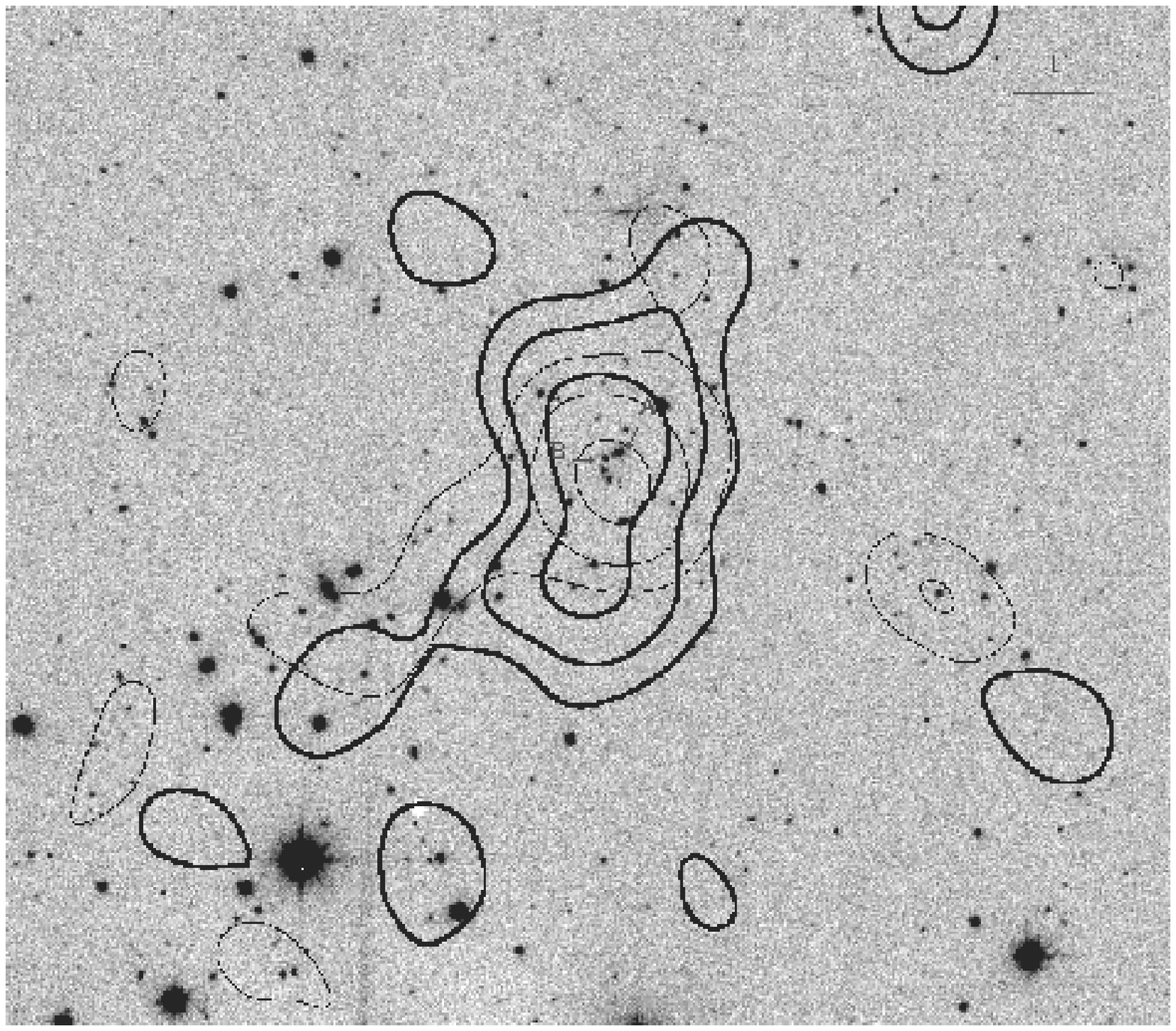,width=6in}}
\figcaption{R-band image obtained with the FLWO 1.2m telescope.
The thin contours are the number density of optical galaxies
(constant linear step of 1.37 per arcmin$^2$) and the thick contours are
the X-ray brightness, smoothed with a 30\arcsec\
FWHM Gaussian (the surface brightness contours are spaced by factors of 2).}
\end{figure}

\begin{figure}
\centerline{\psfig{figure=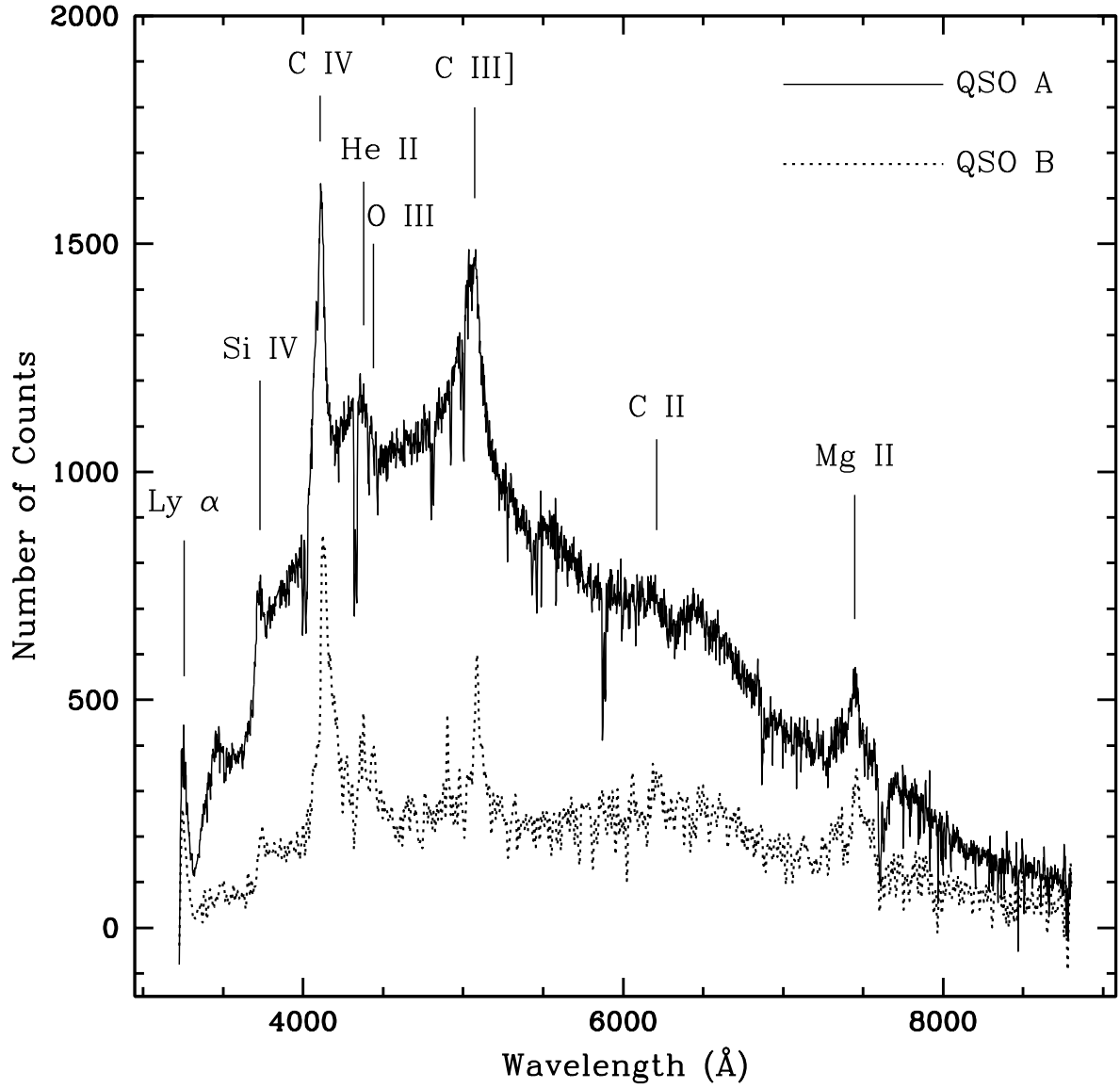,width=6.5in}}
\figcaption{ Spectra of RX\,J0921+4529 A and B. The solid (dashed) line
corresponds to the A (B) component. The B spectrum is multiplied by a factor 2 
for display purposes. Prominent emission lines  are labeled.}
\end{figure}

\begin{figure}
\centerline{\psfig{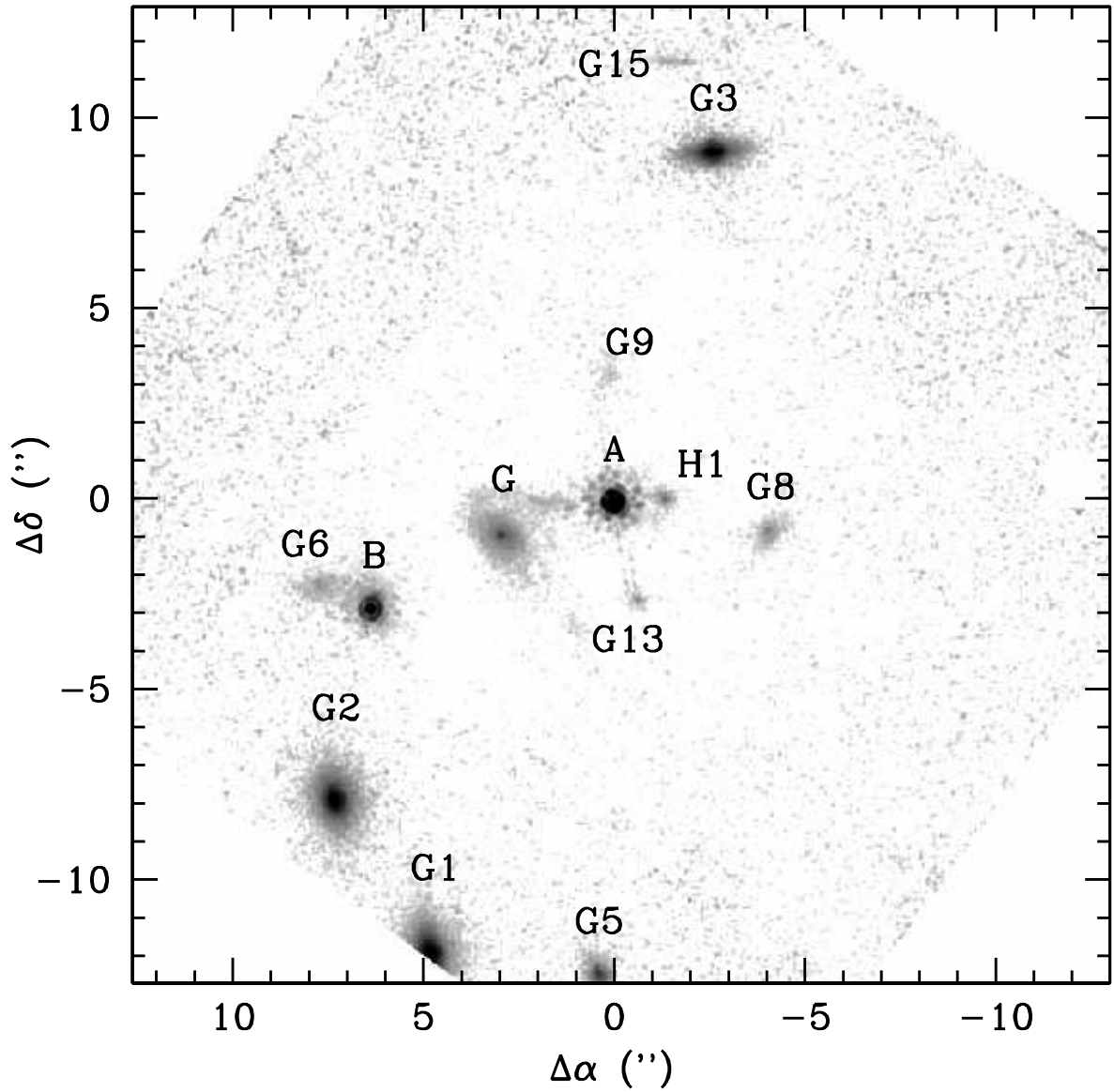}}
\figcaption{The complete NIC2 image 
centered on the A quasar (RA 09h 21m 12.81s, DEC +45\arcdeg\ 29\arcmin\ 04\farcs4 (J2000)).
The two quasar components A and B, the galaxy lens
G and some nearby galaxies are labeled (see Table~2).}
\end{figure}

\begin{figure}
\vspace*{-10cm}
\centerline{\psfig{figure=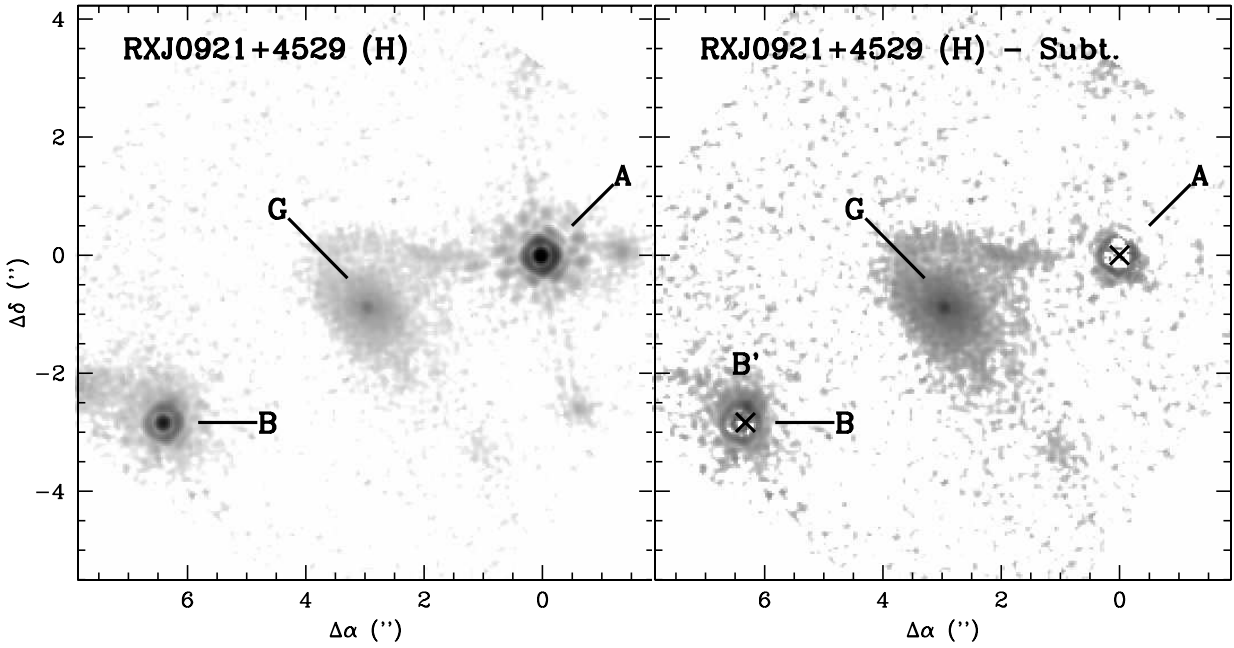,width=8.5in}}
\vspace*{-10cm}
\figcaption{
The left panel shows the HST NIC2 observed image and
the right panel shows the residuals after
quasar subtraction. Quasar images A \& B and  lens galaxy G are labeled.
The B quasar subtraction reveals the B' extended faint object .}
\end{figure}

\begin{figure}
\centerline{\psfig{figure=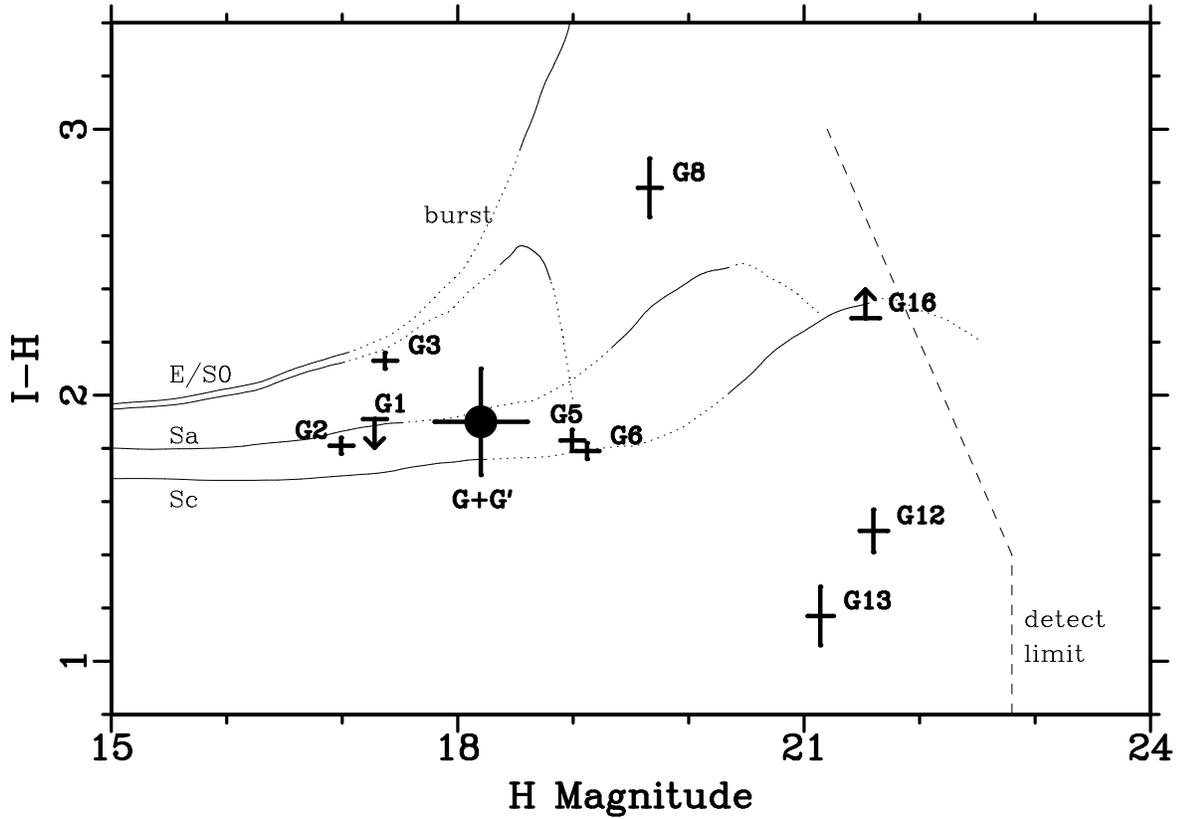,width=7in,angle=-90.}}
\figcaption{HST Colors and total magnitudes for the lens galaxies and their neighbors.
   The curves show photometric evolution models for $L_*$ galaxies,
   alternating between solid and dotted lines
   at intervals of $\Delta z=0.5$\,.
   Galaxies with $L<L_*$ should be shifted to the right of the color-mag curve
   which corresponds to their morphology type.
   The H-band magnitudes have been offset by 0.5\,mag, to correct for the
   SExtractor aperture error (Leh\'ar et~al.\ 2000). 
   The best-fit component photometry for G+G' (see text and Table~1) is shown as a filled circle. 
}
\end{figure}

\begin{figure}
\centerline{\psfig{figure=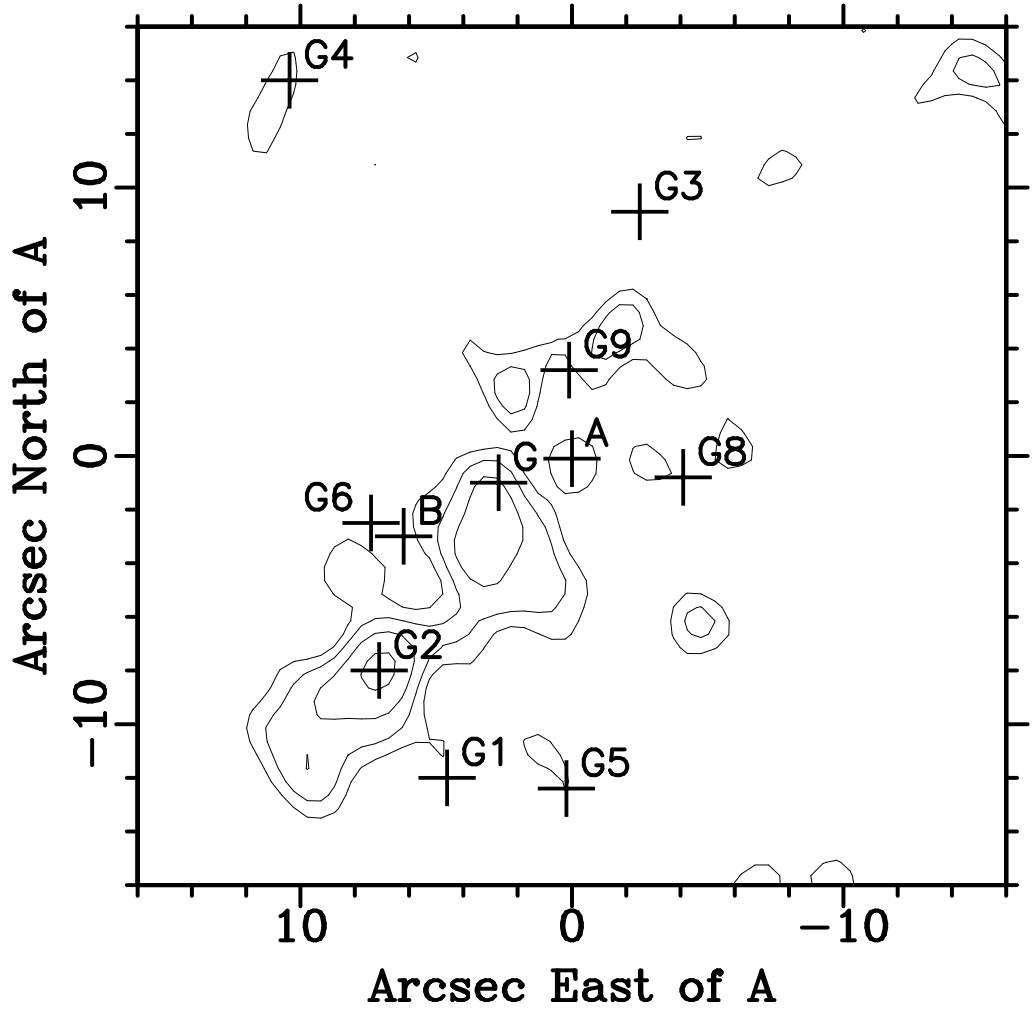,width=6.5in}}
\figcaption{
   VLA C-array map of RX\,J0921+4529 at 3.6\,cm.
   The beam size is approximately $3''$ (FWHM), and
   the radio contours increase by factors of $\sqrt{2}$,
   from twice the off-source rms noise ($46\,\mu\,$Jy/beam).
   Positions of the brightest optical objects are shown. 
   }
\end{figure}

\begin{figure}
\vspace*{-9cm}
\centerline{\psfig{figure=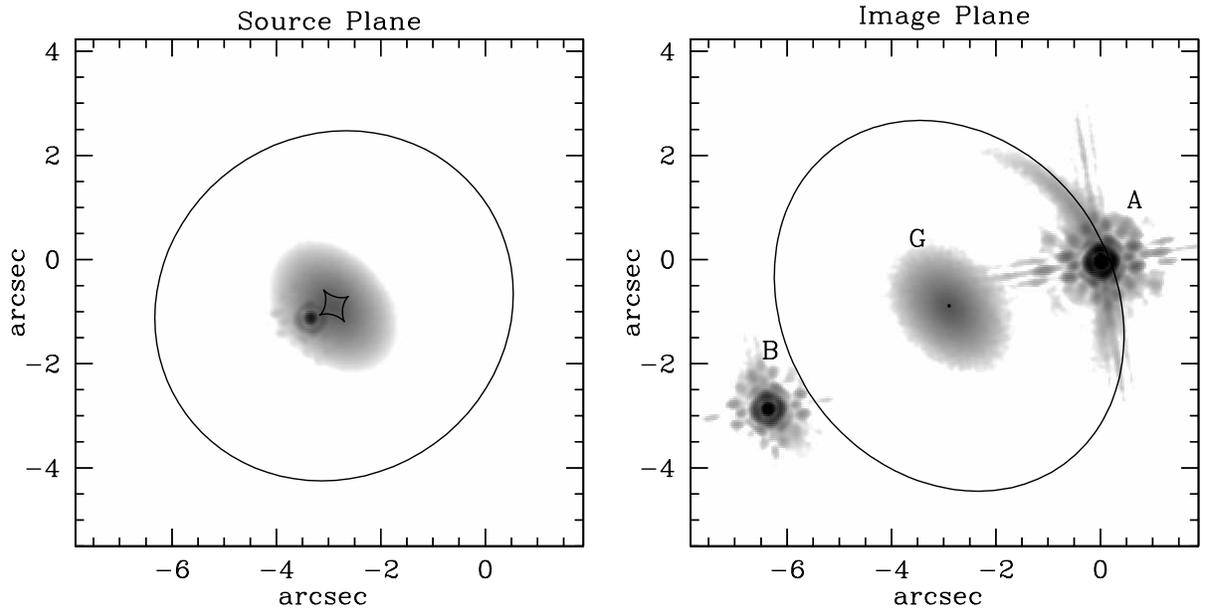,width=7.5in}}
\vspace*{-9cm}
\caption{
The source plane (left) includes an image of the lens galaxy G to
orient the figure, a point source for the quasar and its
host galaxy with a de Vaucouleurs profile.
In the image plane (right) we see the 2 images of the quasar
which exactly fit the observed positions and fluxes. The small
arc from the host galaxy near quasar B has a flux matching that
of B$'$.  However, if B$'$ is a host galaxy, then we should see the
larger, brighter and undetected arc seen near image A.
The caustic and critical lines are shown in the source and image
plane respectively.
}
\end{figure}

\newpage
\begin{deluxetable}{lccccccc}
\small
\tablecaption{HST Astrometry and Photometry of RX\,J0921+4529}
\tablehead{ID  & RA & Dec & Magnitude & Color & $R_e^{(1)}$  & $1-b/a$ & PA  \\
&\arcsec &\arcsec & H & I--H &\arcsec &       &deg}
\startdata
A  &   $\equiv$0     &      $\equiv$0   & 16.90$\pm$0.03 & 1.06$\pm$0.05 & & & \nl
B  & 6.322$\pm$0.005 & --2.837$\pm$0.004 & 18.29$\pm$0.04 & 1.65$\pm$0.05 & & & \nl
B$'$ & 6.22$\pm$0.05   & --2.6$\pm$0.2     & 18.8$\pm$0.7   &  $>2.4$        & 1.0$\pm$0.3   & 0.3$\pm$0.4   & 0$\pm$50 \nl 
G  & 2.908$\pm$0.015 & --0.881$\pm$0.006   & 18.6$\pm$0.2   & 1.6$\pm$0.2   & 0.68$\pm$0.13 & 0.48$\pm$0.09 & 29$\pm$5 \nl
G$'^{(2)}$ & 2.908$\pm$0.015 & --0.881$\pm$0.006   & 19.4$\pm$1.2   & 2.8$\pm$1.7   & 0.5$\pm$0.3 & 0.4$\pm$0.2 & 75$\pm$20\nl
\enddata
\tablecomments{\\
$^{(1)}$ $ R_e$ is the effective radius of the de Vaucouleurs 
profiles for B$'$ and G$'$, and
the scale radius of the exponential disk for G. \\
$^{(2)}$ Due to the complex morphology of the galaxy lens we fit it 
simultaneously with a de Vaucouleurs
(G$'$) and an exponential disk (G) profile to account for as much of the 
galaxy flux as possible. Note that in the text, we refer to 
the flux of G as the sum of the fluxes of these two components.
}
\end{deluxetable}

\begin{deluxetable}{lrrcccrl}
\footnotesize
\tablecaption{Nearby Objects}
\tablehead{ID  & $\Delta$RA & $\Delta$Dec & I & I--H & $\gamma$ & $\theta_\gamma$ & Notes}
\startdata
A    &    0.0 &    0.0 & $18.21\pm0.10$ &  $1.31\pm0.03$ &        &  -73 & z=$1.66$ \\
B    &    6.3 &   -2.9 & $19.91\pm0.10$ &  $1.76\pm0.03$ &        &  -59 & z=$1.66$ \\
G    &    2.7 &   -0.9 & $20.15\pm0.10$ &  $1.60\pm0.03$ &        &      & \\
G1   &    4.6 &  -11.8 & $19.13\pm0.10$ &     $<1.91$    &  0.072 &   -9 & on NIC2 edge \\
G2   &    7.1 &   -7.9 & $19.16\pm0.10$ &  $1.81\pm0.03$ &  0.096 &  -31 & z=$0.3194\pm0.0008$  \\
G3   &   -2.5 &    9.2 & $19.88\pm0.10$ &  $2.13\pm0.03$ &  0.049 &  -28 & z=$0.31\pm0.01$ \\
G4   &   10.4 &   14.1 & $20.91\pm0.10$ &                &  0.021 &   27 & \\
G5   &    0.2 &  -12.3 & $21.32\pm0.10$ &  $1.83\pm0.04$ &  0.025 &   13 & \\
G6   &    7.5 &   -2.4 & $21.41\pm0.10$ &  $1.79\pm0.03$ &  0.058 &  -72 & \\
G7   &   20.4 &    0.1 & $21.83\pm0.10$ &                &  0.013 &   87 & \\
G8   &   -4.1 &   -0.7 & $22.32\pm0.10$ &  $2.78\pm0.11$ &  0.026 &  -88 & \\
G9   &    0.1 &    3.3 & $22.98\pm0.11$ &  $0.66\pm0.14$ &  0.027 &  -33 & \\
G10  &   17.3 &    1.6 & $22.99\pm0.12$ &                &  0.009 &   80 & \\
G11  &   10.5 &  -16.0 & $23.23\pm0.13$ &                &  0.007 &  -27 & \\
G12  &    0.9 &   -3.4 & $23.64\pm0.12$ &  $1.49\pm0.08$ &  0.032 &   39 & \\
G13  &   -0.7 &   -2.6 & $24.18\pm0.13$ &  $1.17\pm0.11$ &  0.020 &   65 & \\
G14  &    8.3 &    6.5 & $24.43\pm0.13$ &     $>3.62$    &  0.008 &   36 & \\
G15  &   -1.4 &   11.6 & $24.75\pm0.15$ &     $>3.66$    &  0.005 &  -19 & \\
G16  &  -14.9 &    0.2 & $24.97\pm0.15$ &     $>2.29$    &  0.003 &  -87 & \\
H1   &   -1.3 &    0.1 &                &     $>7.38$    &  0.047 &  -77 & H=$20.42\pm0.10$ \\
T1   &   54.0 &  -50.7 & $18.32\pm0.10$ &                &  0.016 &  -46 & z=$0.3179 \pm 0.0004$ \\
T2   &   -1.4 &  -30.7 & $18.53\pm0.10$ &                &  0.035 &    8 & \\
T3   &   22.5 &  -22.1 & $18.95\pm0.10$ &                &  0.030 &  -43 & z=$0.309 \pm 0.001$ \\
T4   &   10.9 &  -49.3 & $18.99\pm0.10$ &                &  0.017 &   -9 & z=$0.319 \pm 0.001$ \\
\enddata
\tablecomments{
  SExtractor position offsets from A are given in arcseconds.
  Galaxies on the WFPC2 image within 20\arcsec\ of G are labeled G$x$,
  and those detected only with NIC2 are labeled H$x$.
  Galaxies outside of 20\arcsec\ with large estimated shears are included as T$x$.
  Tidal shear estimates $\gamma$ assume SIS halos 
  with the same redshift and mass-to-light ratio as G,
  and the shear position angles $\theta_\gamma$ are in degrees CCW from North.
   }
\end{deluxetable}

\end{document}